\documentstyle[prl,aps,preprint]{revtex}

\begin{document}

\title {Some exact analytic results for the linear and non-linear\\
response
of atoms in a trap with a model interaction} %
\author{Simon C. Benjamin$^1$,
Luis Quiroga$^2$ and  Neil F. Johnson$^1$} 
\address{$^1$Department of Physics, Oxford University, Oxford, OX1 3PU,
England}
\address {$^2$Departamento de Fisica, Universidad de Los Andes,
Bogota,  A.A. 4976,  Colombia} 
\maketitle

\begin{abstract}

We present an exact expression for the evolution of the wavefunction  of
$N$
interacting atoms in an arbitrarily time-dependent, $d$-dimensional
parabolic
trap potential $\omega(t)$. The interaction
potential
between atoms is taken to be of the form $\xi/r^2$ with $\xi>0$.  For a
constant trap potential $\omega(t)=\omega_0$, we find an exact, infinite
set of
relative mode excitations. These excitations are relevant to the linear
response of the system; they are universal in that their frequencies  are
independent of the initial state of the system (e.g. Bose-Einstein
condensate), the strength $\xi$ of the atom-atom interaction, the
dimensionality $d$ of the trap and the number of atoms $N$. The time
evolution
of the system for general $\omega(t)$ derives entirely from the  solution
to
the corresponding classical 1D single-particle problem. An analytic
expression
for the frequency response of
the $N$-atom cluster is given in terms of $\omega(t)$. We
consider
the important example of a sinusoidally-varying trap perturbation.
 Our treatment, being exact, spans the `linear' and `non-linear' regimes.
Certain features
of the response spectrum are found to be insensitive to interaction
strength
and atom number.

\end{abstract}

\narrowtext

\noindent{\bf I. INTRODUCTION}

Atomic traps have been the subject of much recent research activity, both
theoretical\cite{Burnett} and experimental\cite{Cornell}\cite{Foot}.
 The finite atom number $N$ and reduced effective
dimensionality $d\leq 3$ in atomic traps opens up a fascinating research
area
of confined $N$-atom states. A singularly important milestone in this field has
been the
recent experimental observation of a Bose-Einstein condensate in an
effectively two-dimensional ($d=2$) trap
\cite{Cornell}.  It is to be expected that the application
of time-dependent perturbations to such traps will provide  useful
information
as to  the properties of the $N$-atom system.

According to the preparation of the
atomic trap, the confinement length scales in the three spatial directions
($L_x$, $L_y$ and $L_z$) can, in principle, be quite different yielding
highly
anisotropic traps. Within a simple particle-in-a-box picture the
single-particle
energy level spacing $\Delta E\sim L^{-2}$; it follows that for
$L_x>>L_y,L_z$
the atoms will be stuck in
the lowest $y,z$ subbands  hence freezing out the $y,z$ degrees of freedom.
The anisotropic three-dimensional trap is now effectively
quasi-one-dimensional
in that the atoms only have significant
freedom along the $x$-direction.  Similarly if $L_x\sim L_y >>L_z$ the trap
is effectively
quasi-two-dimensional; if $L_x\sim L_y\sim L_z$ the trap becomes
quasi-three-dimensional.

A variety of theoretical predictions have been made concerning the
properties of atomic traps.
For example, Burnett and co-workers have obtained very interesting results
concerning the linear and non-linear
responses of trapped neutral atoms in a Bose-Einstein condensate by
solving a
time-dependent Gross-Pitaevskii  equation for the condensate
order-parameter or
`wavefunction' \cite{linear}.
Important open questions concern the precise dependence of the $N$-atom
quantum
state on the number of atoms $N$ \cite{Grossmann}, and on the
strength of the interaction between atoms \cite{Kagan}.

Given the general intractability of the $N$-body problem, most
theoretical approaches have been numerical and tend to be based on
macroscopic,
phenomenological models or mean-field approximations.
Our results in this paper complement such studies. In
particular we provide analytic results for the response of an arbitrary
number
$N$  of trapped, neutral atoms by studying a microscopic model Hamiltonian.
We
provide closed-form expressions for the trap properties as a function of
the trap
parameters, thereby yielding insight into the competition between the
confinement and atom-atom interactions. The analytic tractability of our
model
is made possible through a combination of the parabolic form for the
confining
potential in the magnetic traps of interest
\cite{Burnett}, and an inverse-square interaction between atoms, i.e.
$1/r^n$
with $n=2$. The actual short-range repulsive interaction between neutral
atoms
is likely to be better fitted by a larger $n$; however, the
general
features of our results should be qualitatively similar for other
short-range
interactions involving $n>2$.

The paper is organized as follows. In Sec. II we discuss the excitation
spectrum obtained from the time-independent $N$-atom Schrodinger equation
for
an isolated trap. These excitations are relevant for describing the
response
of the trap to a weak, sinusoidal perturbation (i.e. linear response). In
Sec.
III we consider a class of time-dependent perturbations having arbitrary
strength and form, and derive the exact, time-dependent $N$-atom
wavefunction.
As an example,
the exact, non-linear time-dependent response to a sinusoidal perturbation
is developed in Sec. IV; this response is
shown to be largely determined by solutions of the well-known Mathieu
equation. Non-linear features in the system response are identified and
explained. Throughout the paper we will focus on quasi-two-dimensional
traps,
since these systems are attracting much experimental interest. The
generalization of these results to quasi-one-dimensional and isotropic,
three-dimensional traps is straightforward and the corresponding results
are
discussed briefly.  We will also focus on the case of repulsive
interactions
between atoms. The extent to which our results are also applicable  to
attractive atom-atom interactions is discussed in Sec. V, together with the
conclusions.  We will comment throughout on the extent to which our
analytic results can reproduce similar physics to recent numerical work.

\vskip\baselineskip

\noindent{\bf II. ELEMENTARY EXCITATIONS OF AN ISOLATED TRAP}

In this section we focus on the elementary excitations of an isolated trap.
Such excitations govern the linear response of the system in the
presence of a weak, sinusoidal perturbation. We will
specifically obtain an exact, infinite set of relative mode excitations for
the $N$-atom system.

The governing microscopic equation for the energy spectrum of the atom trap
is
the $N$-atom non-relativistic Schrodinger equation with a $d$-dimensional
confining potential $V_C({\bf r})$; this is given by $H\Psi=E\Psi$ with
\begin{equation} H=\sum_i \big[{1\over {2m}}{\bf p}_i^2  + V_C({\bf
r}_i)\big] + \sum_{i<j} V_I({\bf r}_i-{\bf r}_j) \ , \end{equation}  where
the
momentum and position associated with the $i$'th atom are given by ${\bf p}_i$
and ${\bf r}_i$. The magnetic traps of interest have approximately parabolic
confining potentials $V_C$. We will assume that the trap potential for a
$d$-dimensional
trap is a $d$-dimensional isotropic harmonic oscillator, ie.
$V_C=\frac{1}{2}m\omega_0^2r_i^2$. We also assume the atom-atom interaction is
translationally-invariant. The exact functional form of the two-body
interaction $V_I({\bf r}_i-{\bf r}_j)$  between neutral atoms is
unknown; here we focus on a net repulsive interaction between atoms of the
form $\xi(|{\bf r}_i-{\bf r}_j|)^{-n}$ where $n=2$ and $\xi>0$. Note that
for
$n\geq 2$, the interaction is sufficiently `short-range' that we would expect
similar
behavior to emerge for a range of $n$. At the end of the paper we briefly
discuss how our results carry over to an attractive interaction between
atoms
(i.e. $\xi<0$).

We employ standard Jacobi coordinates ${\bf X}_i$  ($i=0,1,\dots ,N-1$)
where   ${\bf X}_0=\frac{1}{N}\sum_j {\bf r}_j$ (center-of-mass),  ${\bf
X}_1=\sqrt{\frac{1}{2}}({\bf r}_2-{\bf r}_1)$,   ${\bf
X}_2=\sqrt{\frac{2}{3}}(\frac{({\bf r}_1+ {\bf r}_2)}{2}-{\bf r}_3)$  etc.
together  with their conjugate momenta ${\bf P}_i$.  The center-of-mass
motion
decouples, $H=H_{\rm CM}({\bf X}_0)+ H_{\rm rel}(\{{\bf X}_{i>0}\})$, hence
$E=E_{\rm CM}+E_{\rm rel}$ and $\Psi=\psi_{\rm CM}\psi_{\rm rel}$. The
exact
eigenstates $\psi_{\rm CM}$ of $H_{\rm CM}$ and eigenenergies $E_{\rm CM}$
are
identical to those of a single particle in a parabolic potential and are
well-known.   We will concentrate
on the relative motion since it is here that effects of atom-atom
correlations
manifest themselves. The non-trivial problem  is to solve the relative
motion
equation  $H_{\rm rel}\psi_{\rm rel}=E_{\rm rel}\psi_{\rm rel}$. We
transform
the relative coordinates  $\{{\bf X}_{i>0}\}$ to standard hyperspherical
coordinates: ${\bf X}_i=r(\prod_{j=i}^{N-2}{\rm cos}\alpha_{j+1}){\rm sin}
\alpha_{i}  e^{i\theta_i}$ with $r\geq 0$ and
$0\leq\alpha_i\leq\frac{\pi}{2}$
($\alpha_{1}=\frac{\pi}{2}$). Physically, the hyperradius $r$ is related to the
root-mean-square atom-atom separation:
\begin{equation}
r^2=\sum_{j=1}^{N-1} X^2_j={1\over N}\sum^{N}_{(i>i')\ i=2}({\bf r}_i-{\bf
r}_{i'})^2 \end{equation}
 As mentioned above, we will focus on $d=2$.
The exact eigenstates of $H_{\rm rel}$ have the form
$\psi_{\rm rel}=R(r)F(\tilde\Omega)$ where $\tilde\Omega$ denotes  the
$(2N-3)$ remaining  $\{\theta,\alpha\}$ variables;
$R(r)$ and $F(\tilde\Omega)$ are  solutions of the hyperradial and
hyperangular  equations respectively.  The
hyperradial
equation is \begin{equation} \big(\frac{d^2}{d r^2}+
\frac{2N-3}{r}\frac{d}{d
r}-\frac{\gamma(\gamma+2N-4)}{r^2}- \frac{r^2}{l_0^4}+\frac{2m E_{\rm
rel}}{\hbar^2} \big)R(r)=0 \end{equation} where
$l_0^2=\hbar(m\omega_0)^{-1}$; the parameter  $\gamma$ is related to
the  eigenvalue of the $\omega_0$-independent hyperangular equation (see
below). Equation (3) can be solved exactly yielding \begin{equation} E_{\rm
rel}=\hbar\omega_0(2n+\gamma+N-1) \end{equation} where $n$ is any positive
integer or zero and  \begin{equation} R_n(r)=\big(\frac{r}{l_0}\big)^\gamma
L_n^{\gamma+N-2} \big(\frac{r^2}{l_0^2}\big) e^{-\frac{r^2}{2l_0^2}}\ \ .
\end{equation}  
where $L$ is the Laguerre Polynomial. Equation (4) provides an exact (and
infinite) set of relative
mode excitations at frequencies $2\omega_0\Delta n$, irrespective of the
initial
eigenstate of the $N$-atom system (e.g. Bose-Einstein condensate). These
are
`breathing' modes; we note that they are very similar in
frequency to the set of modes found to occur near multiples of $2\omega_0$
in
recent numerical calculations  based on the
Gross-Pitaevskii equation \cite{linear}.

The
$\omega_0$-independent hyperangular equation determining $\gamma$ is given by
\begin{equation}
\big[\Theta^2_N+\frac{2m\xi}{\hbar^2}V(\tilde\Omega)\big] F(\tilde\Omega)
=[\gamma(\gamma+2N-4)]F(\tilde\Omega)
\end{equation}
where
\begin{equation}
\Theta^2_N\equiv -\frac{\partial^2}{\partial \alpha_N^2}
+\frac{[2N-6-(2N-4){\rm cos}2\alpha_N]}{{\rm
sin}2\alpha_N}\frac{\partial}{\partial \alpha_N}+{\rm
sec}^2\alpha_N\Theta^2_{N-1}-{\rm
cosec}^2\alpha_N\frac{\partial^2}{\partial\theta_N^2}
\ .
\end{equation}
The quantity $V(\tilde\Omega)$ represents the hyperangular part of the atom-atom
interaction and is given by
\begin{equation}
V(\tilde\Omega)=r^2\sum_{i<j}\frac{1}{|{\bf r}_i-{\bf r}_j|^2} \ \ .
\end{equation}
Since the atom-atom interaction only depends on absolute relative coordinates,
$V(\tilde\Omega)$ commutes with the relative angular
momentum. The relative angular momentum $J$ therefore remains a good quantum
number, hence we  can introduce a further Jacobi transformation
of the relative motion angles $\{\theta_i\}$:  in particular
$\theta'=\frac{1}{N-1}\sum_{i=1}^{N-1}\theta_i$,
$\theta={1\over\sqrt2}(\theta_1-\theta_2)$ etc.
The hyperangular equation now depends on just $(2N-4)$ remaining hyperangles,
i.e. the hyperangles $\tilde\Omega$ excluding $\theta'$. We denote these
$(2N-4)$ hyperangles as $\Omega$. The exact eigenstates of $H_{\rm rel}$ now
have the form $\psi_{\rm rel}=e^{iJ\theta'}R(r)G(\Omega)$  (i.e.
$F(\tilde\Omega)=e^{iJ\theta'}G(\Omega)$).

Unfortunately
the hyperangular equation (Eq. (6)) does not admit complete exact solutions for
$\gamma$ and $F(\tilde\Omega)$, or equivalently $G(\Omega)$. However
some insight  into
the properties of these eigenstates can be gained as follows.
Without loss of
generality, we can choose to rewrite the eigenvalue of the
hyperangular equation (Eq. (6)) in terms of a new quantity $\epsilon$ as
\begin{equation} \epsilon=\frac{\hbar^2}{8}[\gamma(\gamma+2N-4)-
\big(\frac{V_{\rm class}} {\hbar\omega_0}\big)^2-J^2] \end{equation}  where
$V_{\rm class}$ is
the potential energy of the minimum-energy configuration for
{\em classical}
atoms in the trap (i.e. the Wigner solid). This quantity $\epsilon$
is useful in that it isolates purely quantum-mechanical contributions to
$\gamma$, i.e. it does not contain either the classical potential energy of the
atoms
 or their rotational
kinetic energy.
It is straightforward
to show that $V_{\rm class}\propto\xi^{\frac{1}{2}}\omega_0$ and  hence
$\epsilon$ (like $\gamma$) is independent of  $\omega_0$.    We can now recast
the exact relative energy expression as \begin{equation} E_{\rm
rel}=\hbar\omega_0\big[2n+ \big([N-2]^2+J^2+\big(\frac{V_{\rm class}}
{\hbar\omega_0}\big)^2+\frac{8\epsilon} {\hbar^2}\big)^{\frac{1}{2}}
+1\big]\ \ .
\end{equation}
$E_{\rm rel}$ only depends on particle statistics
through
$\epsilon$.  As $\hbar\rightarrow 0$,  $E_{\rm rel}\rightarrow V_{\rm
class}$
and $\epsilon\rightarrow 0$. Physically, $\epsilon$ includes the zero-point
energy in $\Omega$-space   associated with the quantum-mechanical spread of
$G(\Omega)$  about the hyperangles $\Omega$ corresponding to the classical,
minimum energy configuration of the $N$-atom system (Wigner solid).
In the limit of fairly weak interatomic interactions (i.e. small $\xi$), the
spread in $G(\Omega)$ and hence the magnitude of the zero-point energy
$\epsilon$
will be large.  For any given $\xi$, the
spread in $G(\Omega)$ and hence magnitude of $\epsilon$  will also depend on
total
wavefunction
symmetry  requirements; $\epsilon$ will in general be smaller for
a bosonic ground state than for a fermionic ground state due to the lower
number of nodes in $G(\Omega)$ for bosons.

The above analysis which yielded the exact excitation
frequencies $2\omega_0\Delta n$ for quasi-two-dimensional ($d=2$) traps
carries over to quasi-one-dimensional traps ($d=1$) and
isotropic three-dimensional traps ($d=3$). The same breathing-mode
frequencies $2\omega_0\Delta n$ are obtained in each case. For $d=1$ the
quantity $\gamma$ in Eq. (4), and hence the entire energy spectrum
$E_{\rm rel}$, can actually be obtained analytically. This model was first
solved exactly by Calogero \cite{Calogero}. The exact $N$-atom energy
levels for
a quasi-one-dimensional trap are given by $E=E_{\rm CM}+E_{\rm rel}$
with \begin{equation} E_{\rm rel}=\hbar\omega_0 \big( {1\over 2}(N-1) +
{1\over
2}N(N-1)(\tau+{1\over 2}) + k\big)\ \ \end{equation} where $\tau={1\over
2}(1+{4m\xi\over {\hbar^2}})^{1\over 2}$ and $k$ is a positive integer
(N.B.
$k\neq 1$). The hyperradial breathing modes $2\hbar\omega_0\Delta n$ are
included in this relative mode spectrum via the integer $k$ values.
We note that Eq. (11), while exact for non-zero $\xi$, does not yield the
full energy level spectrum for $N$ non-interacting atoms in the limit
$\xi\rightarrow 0$. This feature, which is discussed explicitly in Ref. [7],
results from the singular nature of the inverse-square interaction when applied
in one dimension -- it does not arise in two or three dimensions.

\vskip\baselineskip

\noindent{\bf III. DYNAMICS OF A PERTURBED TRAP}

In this section we extend the above results to include a time-dependent
trap
potential  \begin{equation} \omega^2(t)=\cases{\omega_0^2& for $t\leq 0$\cr
f(t)& for $t>0\ \ $\cr} \end{equation} The treatment will be exact,
analytic
and applicable to any function $f(t)$ with arbitrarily large magnitude,
i.e. we
implicitly include all non-linear effects to all orders.  Specifically, we
will
provide the time-dependent versions of  the hyperradial breathing-mode
wavefunctions given in Eq. (5). These will be found to depend
only
on the solution  of a classical one-dimensional oscillator with trap
potential
$\omega^2(t)$. Again we will specifically consider a quasi-two-dimensional
($d=2$) trap, but the method is equally applicable to both $d=1$ and $d=3$.

The time-dependent Schrodinger equation is given by $H\Psi=i\hbar({\partial
\Psi/\partial t})$. In the presence of the time-dependent trap potential
discussed above, the separation of the center-of-mass and relative motion
is
still exact, $\Psi=\psi_{\rm CM}(t)\psi_{\rm rel}(t)$. It
was
found in Sec. II that the hyperangular equation is independent of
$\omega_0$,
hence those parts of the relative wavefunction $\psi_{\rm rel}$ that
derive
from this equation remain time-independent. The time-dependence of the
total
wavefunction $\Psi$ is therefore only contained in the center-of-mass  and
hyperradial parts. Here  we consider explicitly $\psi_{\rm rel}(t)$ and
hence
the hyperradial part $R(r,t)$. The solution for the trivial center-of-mass
part
$\psi_{\rm CM}(t)$ is exactly analogous.
Following \cite{Camiz} we construct the generating function
\begin{equation}
g(z,r,t)\equiv\sum_{n=0}^\infty R_{n}(r,t) z^n \end{equation} where
$R_{n}(r,t)$
are the solutions of the time-dependent hyperradial equation.  Because the
trap
potential is constant for $t\leq 0$, we can employ Eq. (5) to obtain
\begin{equation} g(z,r,t\leq 0)=\sum_{n=0}^\infty
\big(\frac{r}{l_0}\big)^\gamma
L_n^{\gamma+N-2} \big(\frac{r^2}{l_0^2}\big) e^{-\frac{r^2}{2l_0^2}} z^n
\end{equation} This can be written in closed form using the identity
\begin{equation} \sum_{n=0}^\infty z^n
L^a_n(y)\equiv(1-z)^{-(a+1)}e^{({{zy}\over{z-1}})} \end{equation}
(see Ref.
\cite{tables}) so that  \begin{equation} g(z,r,t\leq 0)=
\big(\frac{r}{l_0}\big)^{\gamma}
 e^{{{z+1}\over{2(z-1)}}(\frac{r}{l_0})^2}(1-z)^{-(a+1)} \end{equation}
where
$a=\gamma+N-2$. We now make the ansatz \begin{equation} g(z,r,t>0)=
\alpha(z,t)
({r\over l_0})^{\gamma} e^{\alpha'(z,t) r^2} \end{equation} which can be
shown to satisfy
the time-dependent hyperradial equation \begin{equation} \big(\frac{d^2}{d
r^2}+ \frac{2N-3}{r}\frac{d}{d r}-\frac{\gamma(\gamma+2N-4)}{r^2}-
({{m}\over\hbar})^2\omega^2(t)r^2)g(z,r,t)=-{{2im}\over\hbar}
\frac{\partial}{\partial t}g(z,r,t) \end{equation}  and the $t=0$ boundary
condition in Eq. (16), provided \begin{equation}
\alpha(z,t)=[\eta(t)]^{-(a+1)}exp[2i\theta(t)(a+1)](1-z\
exp[2i\theta(t)])^{-(a+1)} \end{equation} and \begin{equation}
\alpha'(z,t)={{im}\over{2\hbar}}({{\eta\dot(t)}\over{\eta(t)}}
-2i\theta\dot(t)(1-z\ exp[2i\theta(t)])^{-1}) \end{equation} where
$\eta(t)=|\eta(t)|e^{i\theta(t)}$ solves the {\em classical one-dimensional
oscillator} \begin {equation} \eta\ddot(t)+f(t)\eta(t)=0 \end {equation}
with
boundary conditions $\eta(0)=1$ and $\eta\dot(0)=-i\omega_0$. We may then
expand
$g(z,r,t)$ using the identity (15), and compare coefficients of $z^n$ with
the
defining Eq. (13) to obtain the desired time-dependent  wavefunctions
(unnormalised): 
\begin{equation} 
R_n(r, t>0)=
|\eta(t)|^{1-N} y^\gamma
exp[i(
\theta(t)(2n+a+1)+{{y^2}\over{4\omega_0}}{d\over{dt}}|\eta(t)|^2
)]  e^{-{1\over2}y^2}L_n^a(y^2)\ . 
\end{equation} 
where $y\equiv{r\over{|\eta|\l_0}}$. In the static case this expression reduces
to equation (5). If for a particular $f(t)$ we
are able to solve Eq. (21), then Eq. (22) provides a complete description
for the evolution of the initially stationary hyperradial state $R_n$ as
the
trap becomes time-dependent for $t>0$. As noted previously, since the
remaining hyperangular part of $\psi_{\rm rel}$ is time-independent, Eq.
(22)
together with the (exactly analogous) expression for the center-of-mass
entirely determines the time-evolution of the total wavefunction
$\Psi=\psi_{\rm CM}(t)\psi_{\rm rel}(t)$.

We now comment on the significance of our results,
particularly with respect to the Bose-condensed atomic gas. Our results for the
breathing modes in Sec. II and for the dynamical response in Sec. III are valid
for all strengths of the atom-atom interaction ($\xi$), all trap sizes
($\omega_0$) and all numbers of atoms ($N$). They are independent of the precise
form of the initial state of the system and are therefore true in both the
Bose-condensed and non-condensed regimes. Since we do not know $\gamma$ (or
$F(\tilde\Omega)$) explicitly, our model does not allow us to predict a
temperature at which the gas will be substantially Bose-condensed. This is
consistent with the expectation that the condensation temperature has a
complicated dependence on $\omega_0$, $\xi$ and $N$.

In deriving the time-dependent
wavefunction of Sec. III, we took the $N$-atom system to be in
a single eigenstate of the time-independent Schrodinger equation
up until the moment the
perturbation is turned on (i.e. for $t<0$). The linearity of the
Schrodinger equation implies that any initial wavefunction consisting of a
sum of eigenstates will evolve as the sum of its parts, and may therefore be
written as a sum of terms such as Eq. (22). In Sec. IV below, the
calculated response spectrum is also
derived under the assumption of a single eigenstate for
$t<0$, however the initial state can be generalized to a sum of
eigenstates {\em without altering the response spectrum}. The results we present
in this paper are therefore remarkably general in their applicability.

\vskip\baselineskip

\noindent{\bf IV. EXAMPLE: SINUSOIDAL PERTURBATION}

As an illustration of the applicability of our formalism,
we now take a particular $f(t)$ and
employ Eq.
(22) to
calculate  the frequency response spectrum of the system. We find that
the spectrum exhibits some features  which are essentially independent of
$\gamma$. Since the
$\xi$-dependence only enters the response through
$\gamma$, it follows that any observables found to be substantially
independent of $\gamma$ are generic to all atom traps {\em independent} of
the
strength of the atom-atom interaction $\xi$.

We consider
the
specific case of a sinusoidal perturbation turned on at $t=0$:
\begin{equation}
\omega^2(t)=\cases{\omega_0^2& for $t\leq 0$\cr
f(t)=\omega^2_0-\omega_1^2(1-{\rm cos}(2\Omega t))& for $t>0\ \ $\cr}
\end{equation}
A sketch of this function appears at the top of Fig. 1. It would be
quite possible to produce this trap perturbation experimentally.
The solution to the
classical equation (Eq. (21)) is a general Mathieu function, which may be
written in the form
\begin{equation}
\eta(t)=A e^{\mu\Omega t}\sum_{n=-\infty}^{\infty} c_{2n} e^{2in\Omega t}+
        B e^{-\mu\Omega t}\sum_{n=-\infty}^{\infty} c_{2n} e^{-2in\Omega t}
\end{equation} where $\mu$ and $\{c_{2n}\}$ are determined by a set of
simultaneous
equations (see Ref. \cite{tables}).

In the main part of Fig. 1  we show the behavior of $\mu$ as a
function of the parameters
$\omega_0$, $\omega_1$ and $\Omega$. This behavior is non-trivial; in
certain regions (shown white) of the
parameter space $\mu$ is purely imaginary while in others (shown
dark) it has  a real part. We define $\beta\equiv Im\{\mu\}$. A non-zero real
part of
$\mu$ indicates that the classical particle is resonating with
the oscillating trap; the particle's  oscillations then become infinitely
large as $t\rightarrow\infty$. The corresponding effect  on the quantum
mechanical system, which depends on time only through $\eta(t)$, will  be
an
increase in energy and a decrease in localization of $R(r,t)$ and hence
$\psi_{\rm rel}$. This spreading in $\psi_{\rm rel}$ implies an increase in
the average atom-atom separation, and will lead to atoms escaping from any
realistic trap having a finite depth.
 For small $\omega_1$,  $\mu$ has a real component only when the perturbing
frequency $\Omega$ is equal to  an integer {\em fraction} ${1\over n}$ of
the
trap potential $\omega_0$. We emphasize that our treatment is
exact for any amplitude
$\omega_1^2$ of the sinusoidal perturbation. Indeed the theory remains
sound even when $\omega_1^2>{1\over2}\omega_0^2$, in which case the trap
becomes repulsive for part of each oscillation.

In the remainder of this section we concern ourselves with solutions which
are
`stable', i.e. points in the parameter space corresponding to a purely
imaginary $\mu$.
As in the numerical studies of
Ref. \cite{linear}, we will calculate the frequency response spectrum of the
atomic gas by Fourier analyzing the temporal variation of the
 gas density, the `single-particle-density', at a fixed point in space. We
define the single-particle-density at a point ${\bf S}$ as
\begin{equation}
d({\bf S},t)=\sum_{i=1}^N\int|\Psi
({\bf r}_1,{\bf r}_2,...,{\bf r}_{i-1},{\bf S},{\bf r}_{i+1},...{\bf r}_n;t)|^2
\prod_{j\neq i}d{\bf r}_j
\end{equation}
where $\Psi$ is the total wavefunction at time $t$ which, in our case, exactly
separates into center-of-mass,
hyperradial and angular parts. If we set $\bf{S=0}$ (i.e. the trap center), we
find
\begin{equation}
d({\bf S=0},t)=N\int
|\psi_{CM}^{\bf S=0}(t)|^2|R_n(r,t)|^2|F(\tilde\Omega)|^2 
r^{2N-3}dr d\tilde\Omega
\end{equation}
The superscript on $\psi_{CM}(t)$ indicates that
under the ${\bf S=0}$ constraint, the center-of-mass coordinate is no longer
independent of the other coordinates. Examining the form of $\psi_{CM}$ and $R$
we find that the integral in Eq. (26) may be rewritten in time-independent form
using the scaled variable
$y={r\over{|\eta(t)|l_0}}$, as employed in Eq. (22).
This transformation to a time-independent integral results in an external factor
$|\eta(t)|^{-2}$, thus the density is found to have a simple time-dependence:
\begin {equation}
d({\bf S=0},t)\propto |\eta(t)|^{-2}\ .
\end{equation}
Interestingly, this expression (Eq. (27)) is completely general for {\em any}
time-varying trap  $\omega^2(t)$; here we have chosen to consider a
sinusoidal form for which the
corresponding $\eta(t)$ may be found analytically.
The expression for $d({\bf S=0},t)$ is independent of both $N$ and the unknown
quantity
$\gamma$, thus the peak positions in the Fourier
transform must be
independent of variations either in the atom-atom interaction strength
$\xi$ or
in the number of atoms $N$ confined in the trap. Figure 2 displays the
Fourier transform of Eq.
(27)
calculated with a particular choice of the parameters
$\omega_0$ and $\omega_1$.
We see a number of sharply  defined peaks;
the frequencies at which these peaks occur are all of the form
$n\Omega+m\Omega\beta$ where $n$ and $m$ are integers. The value of $\beta$
may
be found from Fig. 1; for the present choice of parameters $\beta=1.409$.

Figure 1 also shows that when the perturbation amplitude
$\omega_1$ is small then  $\beta\approx{\bar{\omega}\over \Omega}$,
where $\bar{\omega}^2=\omega_0^2-\omega_1^2$ is the mean
trap confinement. Thus for a
weak-to-moderate perturbation, the peaks in the response lie at sums and
differences  of $\bar{\omega}$ and the driving frequency
$\Omega$; i.e. at $n\Omega+m\bar{\omega}$. Figure 1 shows us exactly how
$\beta$ (and hence the spectrum) changes as we move to the strongly
non-linear regime  ($\omega_1\approx\bar{\omega}$). The parameters chosen
for Fig. 2 correspond to a `moderate' perturbation
($w_1^2=0.138\omega_0^2$); the peaks marked $n\beta$ in Fig. 2 deviate
from
$n\omega_0$ by about $8\%$, and from $n\bar{\omega}$ by about $0.3\%$ .
Although a direct comparison with the non-linear
response of Ref. \cite{linear} is not practical
because
of the different form of the perturbation (i.e.
white noise) it is interesting to note that the same types of non-linear
phenomena
are observed; in particular, harmonic generation and frequency mixing. As
noted in Ref. \cite{linear}, such non-linear effects open up the
possibility of
non-linear atom optics based on coherent matter waves.

\vskip\baselineskip

\noindent{\bf V. CONCLUSION}

We have presented an analytically-solvable model of the
quantum-mechanical time-evolution of $N$ interacting atoms
in an arbitrarily time-dependent, $d$-dimensional parabolic
trap potential $\omega(t)$. The solution allows us to determine the
resonance frequencies of the system in response to both weak (linear
response) and strong (non-linear response) perturbations. An exact,
infinite set
of relative mode excitations were found for a constant trap potential
$\omega(t)=\omega_0$; these excitations are universal in that their
frequencies
are independent of the initial state of the system (e.g. Bose-Einstein
condensate), the strength $\xi$ of the atom-atom interaction, the
dimensionality
$d$ of the trap and the number of atoms $N$. The specific example of a
sinusoidally-varying trap perturbation was employed to demonstrate the
formalism. Certain features in the system response were found to be
insensitive
to interaction strength and atom number.

Finally we will discuss the extent to which our model can describe an
attractive interaction between atoms (i.e. $\xi<0$). The general formalism
using
hyperspherical coordinates is still valid; the difference is that
$\epsilon$ (see Eq. (9)) can now be negative, hence there is a possibility
of
collapse of the $N$-atom system if the attraction is too strong (i.e. the
atoms
collapse to form an infinitely dense gas). Such a complete collapse is an
artefact of the inverse-square attractive  interaction. It could be
prevented by
inserting an additional hard-core repulsion into the model, but the
solvability
of the model would be lost.
Consider first a pair of atoms ($N=2$) with an attractive interaction
($\xi<0$). In a quasi-one-dimensional trap ($d=1$), it can easily be shown
that
two-atom collapse occurs if the strength of the attractive interaction
$|\xi|>\frac{\hbar^2}{4m}$; only weak attractive interactions  where
$|\xi|<\frac{\hbar^2}{4m}$ are therefore described by the present theory.
The
reason that a pair of atoms with a weak attraction can support a
non-collapsed
ground state in the trap is that the negative potential energy is offset by
a
large, positive kinetic energy as a result of the uncertainty principle.
 In an isotropic three-dimensional trap ($d=3$), two-atom
collapse will occur if the strength of the attractive interaction
$|\xi|>{\frac{\hbar^2}{4m}}(1+4l(l+1))$ where $l$ is the relative angular
momentum of the two atoms. Hence for $l=0$, the same condition holds as for
$d=1$.  For quasi-two-dimensional traps, it turns out that two-atom
collapse
will occur for $|\xi|>{\frac{\hbar^2 l_z^2}{m}}$ where $l_z$ is the
component of
relative angular momentum perpendicular to the two-dimensional plane.
Therefore,
for $l_z=0$ an arbitrarily small attractive potential will cause two-atom
collapse. For $N$ atoms in a quasi-one-dimensional trap, the same condition
holds as for two atoms. For quasi-two and three-dimensional traps, the
exact
condition is not known since the problem is not completely solvable
($\gamma$
and hence $\epsilon$ are not known exactly for $N>2$). It remains to be
seen,
therefore, whether the increase in kinetic energy during collapse is
sufficient
to offset the decrease in potential energy. If
such a condition were to be satisfied, the $N$-atom system with attractive
interactions would form a ground state that was qualitatively
different from the $N$-atom system with repulsive interactions. As a matter
of
interest, we note that a suggestion of possible alternative states for
$N$ atoms with attractive interactions ($^7$Li) has recently
appeared \cite{Physics}.

We would like to thank Keith Burnett, Peter Ruprecht and Nikos Nicopoulos
for
very useful discussions. S.C.B. is supported by an EPSRC studentship.
Partial
funding was also provided by COLCIENCIAS under Project no. 1204-05-264-94.

\newpage \centerline{\bf Figure Captions}

\bigskip

\noindent Figure 1. Top: The trap potential $\omega^2(t)$ (cf. Eq.(23)).
Bottom: Contour plot showing the behavior of the constant $\mu$
in
the Mathieu function $\eta(t)$ (Eq. (21)) as a function of the
trap potential parameters. This quantity $\mu$ is complex,
$\mu=Re\{\mu\}+i\beta$.
White regions: $Re\{\mu\}=0$ and the contours are lines of
constant $\beta$; for $\omega_1\rightarrow0$,
$\beta\rightarrow{\bar{\omega}\over\Omega}$. Dark regions:
$Re\{\mu\}>0$ and the contours are lines of constant $Re\{\mu\}$ in
increments of 0.2.

\bigskip

\noindent Figure 2. Frequency response spectrum showing Fourier
amplitude as a function of frequency response for a trap with parameters
$\omega_1^2=0.138\omega^2_0$ and $\Omega^2=0.431\omega^2_0$.  For
these parameters
$\beta=1.41$; the $\beta$ for any particular trap parameters
($\omega_0$,$\omega_1$,$\Omega$) may be found using Fig. 1. Note that
the two central lines (shown as cut-off)
exceed the height of the graph by a factor of 3.

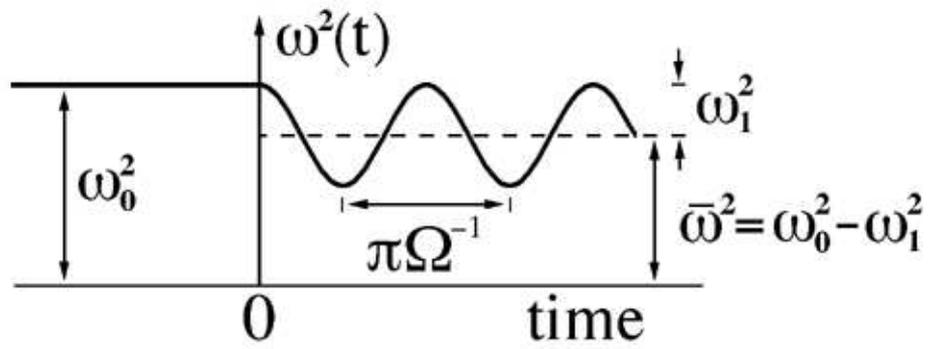
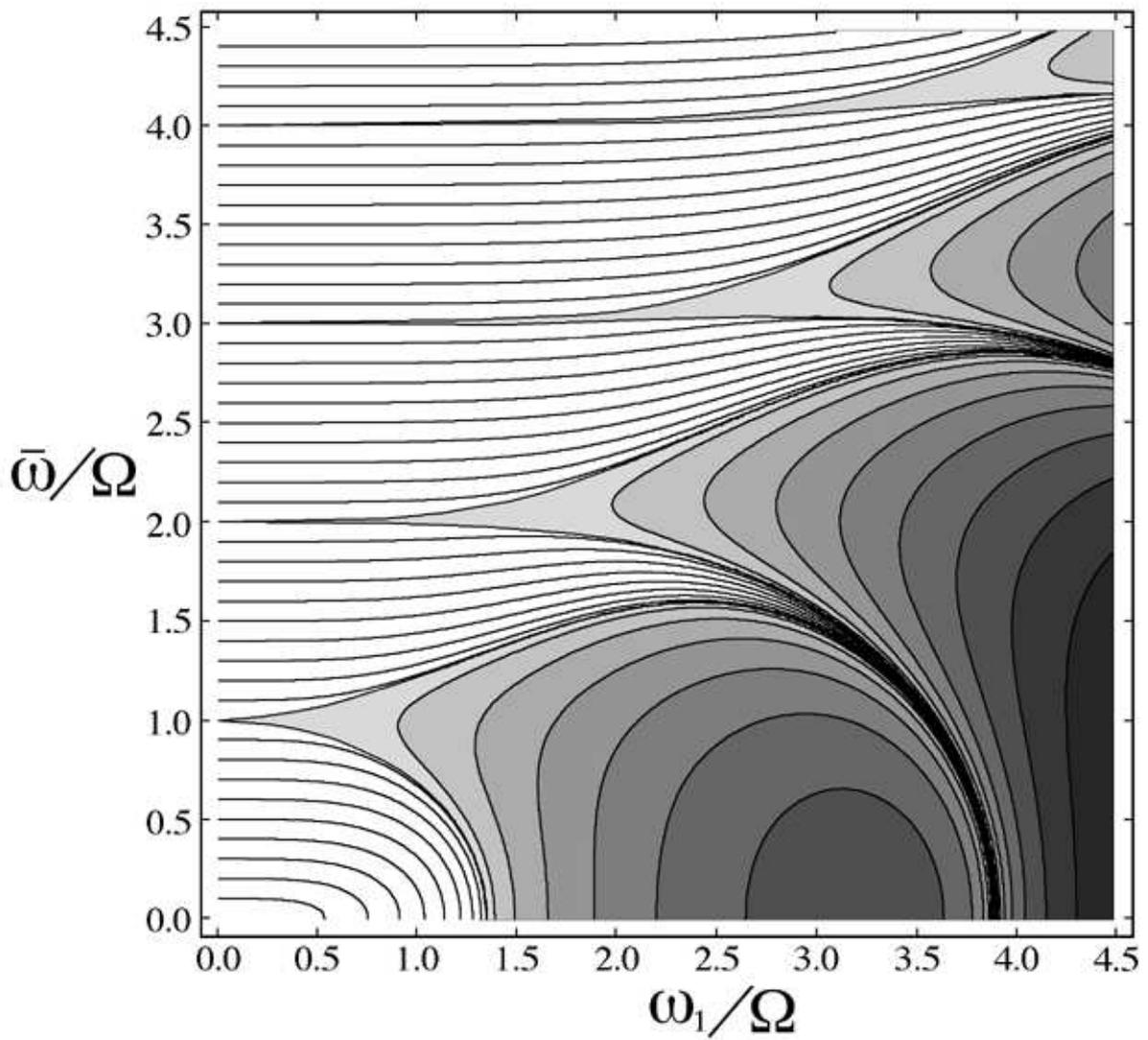
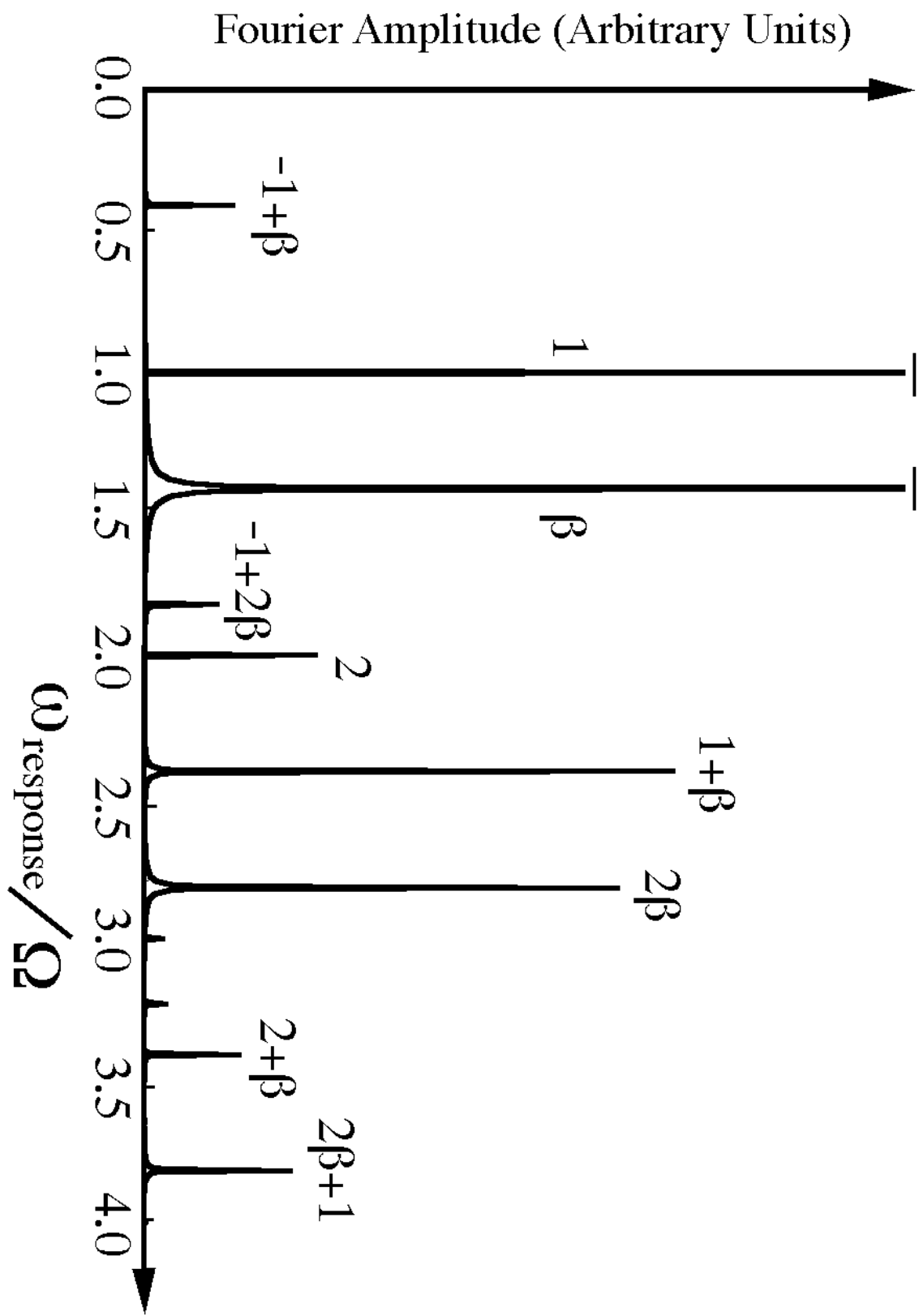
\end{document}